\documentclass[
]{ceurart}

\usepackage{listings}
\begin{document}

\copyrightyear{2026}
\copyrightclause{Copyright for this paper by its authors.
  Use permitted under Creative Commons License Attribution 4.0
  International (CC BY 4.0).}

\conference{GenAIECommerce'26: The Third Workshop on Agentic and Generative AI for E-Commerce, co-located with RecSys, September 28, 2026, Minneapolis, MN, USA}

\title{Beyond Ranking Accuracy: Evaluating LLM-Cited Feature Rationales for Next Basket Repurchase Recommendation}

% \author{Anonymous Author(s)}
% TODO(author): replace placeholder author block with final author list.
\author[1]{Yanan Cao}[%
  % orcid=0000-0000-0000-0000,
  email=yanan.cao@walmart.com,
]
% \cormark[1]
\address[1]{Walmart Global Tech, Sunnyvale, CA, United States}

\author[1]{Anay Dombe}[%
  % orcid=0000-0000-0000-0000,
  email=anay.dombe@walmart.com,
]
% \cormark[1]
% \address[1]{Walmart Global Tech, Sunnyvale, CA, United States}

\author[1]{Murali Mohana Krishna Dandu}[%
  % orcid=0000-0000-0000-0000,
  email=murali.dandu@walmart.com,
]
% \cormark[1]
% \address[1]{Walmart Global Tech, Sunnyvale, CA, United States}

\author[1]{Shreeranjani Srirangamsridharan}[%
  % orcid=0000-0000-0000-0000,
  email=shreeranjani.srirang@walmart.com,
]
% \cormark[1]
% \address[1]{Walmart Global Tech, Sunnyvale, CA, United States}

\author[1]{Sinduja Subramaniam}[%
  % orcid=0000-0000-0000-0000,
  email=sinduja.subramaniam@walmart.com,
]
% \cormark[1]
% \address[1]{Walmart Global Tech, Sunnyvale, CA, United States}

\author[1]{Yogananth Mahalingam}[%
  % orcid=0000-0000-0000-0000,
  email=yogananth.mahalingam@walmart.com,
]
% \cormark[1]
% \address[1]{Walmart Global Tech, Sunnyvale, CA, United States}

\author[1]{Evren Korpeoglu}[%
  % orcid=0000-0000-0000-0000,
  email=ekorpeoglu@walmart.com,
]
% \cormark[1]
% \address[1]{Walmart Global Tech, Sunnyvale, CA, United States}

\author[1]{Kannan Achan}[%
  % orcid=0000-0000-0000-0000,
  email=kannan.achan@walmart.com,
]
% \cormark[1]
% \address[1]{Walmart Global Tech, Sunnyvale, CA, United States}

% \cortext[1]{Corresponding author.}
\begin{abstract}
Next-basket repurchase recommendation is commonly formulated as a ranking task: given a customer's purchase history, the system ranks previously purchased items that may be needed again. In production settings, however, ranking accuracy is only one component of recommendation quality. Customers may also benefit from concise evidence about why an item is recommended now. Large language models (LLMs) offer a potential mechanism for surfacing such evidence by producing feature-based, human-readable rationales grounded in interpretable behavioral signals rather than ranking scores alone.
We construct interpretable repurchase features spanning cadence, frequency, recency, user behavior, and item popularity, and evaluate LLMs on two public grocery datasets and one proprietary retail dataset. We investigate (1) whether off-the-shelf LLMs can use these features as next-basket scorers relative to personal-frequency and supervised rankers, and (2) whether the features cited by LLMs as rationales carry outcome-grounded ranking signal. For the latter, we compare LLM-cited features with model-specific attribution methods under a cross-model feature-masking protocol that measures ranking-quality degradation after masking selected features.
Our results provide a nuanced view of LLM scorers for repurchase recommendation. LLM scores are not competitive with supervised rankers, suggesting that off-the-shelf LLMs should not be used directly as standalone repurchase recommenders. However, changes in prompt and evidence representation can yield stronger outcome-grounded feature-masking results in some settings even when ranking performance does not improve; this pattern is dataset-dependent and does not consistently match model-specific attribution baselines. These findings suggest a practical role for LLMs as validated explanation components: not as primary next-basket rankers, but as tools for surfacing behavioral evidence whose outcome-grounded relevance should be evaluated separately from ranking accuracy.

\end{abstract}


\maketitle

\section{Introduction}

Next-basket repurchase recommendation is commonly formulated as a ranking task: given a customer's purchase history, the system ranks previously purchased items that may be needed again. In user-facing settings, however, ranking accuracy is only one aspect of recommendation quality. A customer may see a recommended item but still need to recall when it was last purchased, how often it is usually repurchased, and whether it may be needed again now. Without such evidence, the customer must reconstruct the rationale behind the recommendation.

This gap is particularly relevant for repurchase recommendation, where useful signals are often observable and behaviorally interpretable. A candidate item may be supported by purchase frequency, recency, cadence, inter-purchase intervals, user shopping patterns, or global reorder behavior. Although supervised recommenders can use these features to rank items, their scores do not reveal which signals support a specific recommendation. Large language models (LLMs) offer a potential way to translate structured behavioral evidence into concise rationales, such as ``You bought this item in 4 of your last 6 orders.'' However, fluent explanations may still cite features that are plausible but not relevant to future repurchase outcomes.

\begin{figure}[h]
  \centering
  \includegraphics[width=0.85\textwidth]{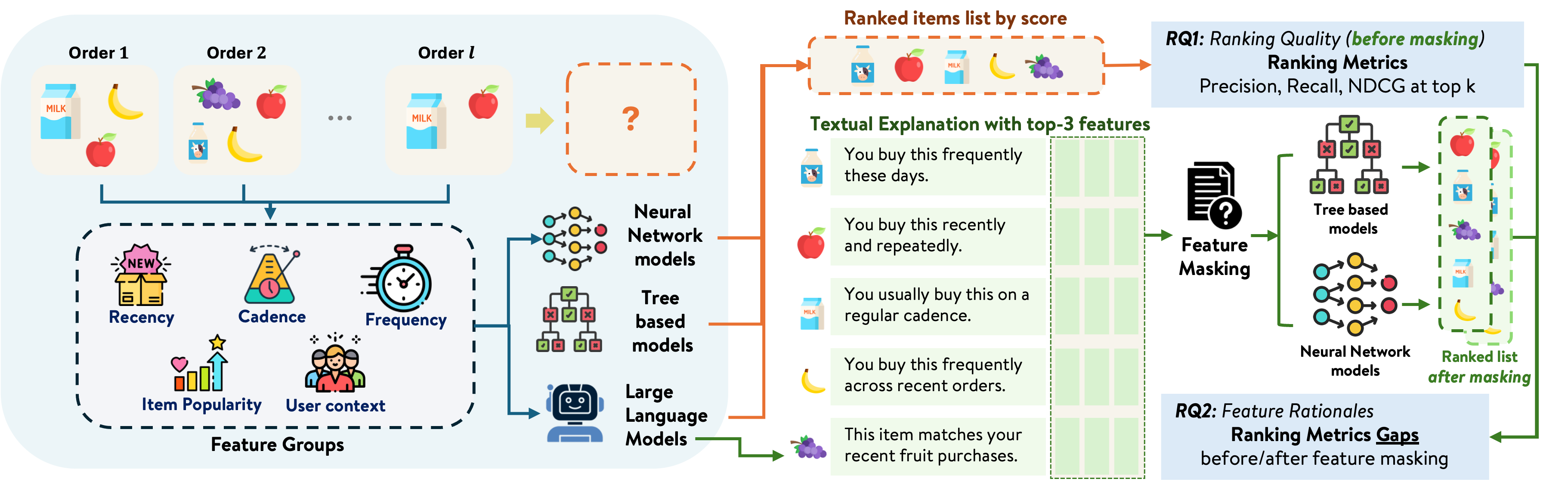}
    \caption{Overview of the experimental framework. Historical purchase sequences are converted into structured behavioral features used by both supervised rankers and LLM scorers. We evaluate LLM score utility in RQ1 using ranking metrics, and feature-rationale quality in RQ2 using cross-model feature masking and ranking-metric gaps.}
  \label{fig:overview}
\end{figure}

In this work, we study LLM feature rationales for next-basket repurchase recommendation beyond ranking accuracy. We first evaluate whether an LLM scorer can extract useful recommendation signal from the same structured behavioral features used by supervised machine learning rankers. We then focus on the central question of this paper: whether the features cited by the LLM scorer capture behavioral signals that remain important across different recommendation model classes. This cross-model perspective is important because production recommenders may involve different architectures, ensembles, or evolving model pipelines, and a useful explanation should not merely mimic the attribution pattern of one specific model. If LLM-cited features remain important across target models, this suggests that the LLM is capturing more general signals associated with future repurchase behavior. To answer this, we compare LLM-cited features with model-specific attribution methods for two parallel target models: TreeSHAP for a tree-based ranker and integrated gradients for a vanilla neural network (VNN). We evaluate all feature selectors through the same cross-model feature-masking protocol.

We organize the study around two research questions:

\begin{description}
  \item[RQ1.] Can an off-the-shelf LLM use structured repurchase features as a next-basket scorer compared with heuristic and supervised rankers?

  \item[RQ2.] Do LLM-cited feature rationales capture outcome-relevant behavioral signals across different recommendation model classes?
\end{description}

For RQ2, we compare LLM-cited features with attribution methods from both target model families and measure the ranking degradation caused by masking the selected features. We use \emph{outcome-grounded} in this operational sense: cited features are considered more outcome-grounded when masking them produces greater degradation in ranking metrics evaluated against future repurchase labels. This criterion measures outcome-relevant ranking signal rather than causal relevance or complete explanation faithfulness.

Our contributions are summarized as follows:
\begin{itemize}
  \item We formulate next-basket repurchase explanation as a feature-rationale evaluation problem, where LLMs are asked not only to assign recommendation scores but also to cite explicit behavioral features such as cadence, frequency, recency, user context, and item popularity.

  \item We evaluate off-the-shelf LLMs as standalone feature-based scorers across two public grocery datasets and one proprietary retail dataset, and show that structured behavioral evidence alone is insufficient for LLMs to replace supervised next-basket rankers.

  \item We propose a cross-model feature-masking evaluation protocol for LLM-cited rationales, comparing them with TreeSHAP and integrated-gradient attributions across tree-based and neural target models. The results show that ranking quality is not sufficient to characterize rationale quality: the two evaluation axes can diverge under changes in prompt and evidence representation.

  \item We complement the feature-masking evaluation with a text-to-evidence groundedness check on Instacart, distinguishing whether cited evidence is outcome-grounded from whether generated text accurately reflects that evidence.
\end{itemize}

\section{Related Work}

\subsection{Explainable Recommendation}

Explainable recommendation studies how recommender systems can provide rationales that help users understand why an item is recommended~\cite{zhang2018explainable}. Prior work has considered explanations based on reviews, item attributes, collaborative signals, knowledge graphs, and user-item interaction patterns. In repurchase recommendation, however, useful explanations often depend on structured behavioral evidence rather than textual preference alone. A customer-facing rationale may need to refer to purchase frequency, recency, cadence, inter-purchase interval, or reorder behavior, all of which have also been used as predictive signals in recent next-basket repurchase modeling~\cite{cao2026case}. Our work focuses on this setting: explaining next-basket repurchase recommendations using structured temporal and behavioral evidence.

\subsection{LLMs for Recommendation Explanations}

Recent work has explored large language models (LLMs) as explanation generators for recommender systems. XRec uses LLMs to provide explanations for user behavior in recommender systems by incorporating collaborative signals~\cite{xrec2024}. LLMXRec proposes a two-stage framework in which a recommender model and an LLM-based explanation generator work together to produce personalized recommendation explanations~\cite{llmxrec2023}. These works show that LLMs can improve the fluency and personalization of recommendation explanations. Our focus is different: rather than evaluating explanation text alone, we study whether the features cited by an LLM as recommendation rationales correspond to behavioral signals that remain important across different trained rankers.

\subsection{LLMs as Interfaces to Existing Feature Attributions}

Prior work has used LLMs to make existing machine-learning explanations more accessible. Conversational XAI systems such as TalkToModel support natural-language interaction with model explanations~\cite{slack2023talktomodel}, while other work translates feature-importance or SHAP-style outputs into natural-language responses~\cite{tekkesinoglu2024feature,zeng2024shap}. In these settings, the LLM primarily verbalizes explanations produced by an external attribution method.

Our setting is different. The LLM selects feature rationales directly from a structured behavioral evidence card without access to model parameters, gradients, tree structures, or attribution values. Our goal is not to replace model-specific attribution methods for explaining a particular trained ranker; rather, we ask whether independently selected behavioral evidence remains relevant across different model classes. TreeSHAP and integrated gradients therefore serve as comparison and validation mechanisms rather than targets that the LLM is expected to reproduce or outperform.

\subsection{Model-Specific Attribution and Rationale Grounding}

Feature attribution methods assign importance to model inputs, including SHAP and TreeSHAP for tree-based models~\cite{lundberg2017shap,lundberg2020trees} and integrated gradients for differentiable neural models~\cite{sundararajan2017axiomatic}. Prior work on explanation faithfulness further distinguishes explanations that appear plausible from those grounded in the evidence or behavior underlying a prediction~\cite{jacovi2020faithfully}, while rationale-evaluation methods study whether selected evidence is sufficient or necessary for model outputs~\cite{lei2016rationalizing,deyoung2020eraser}.

We adapt this perspective to tabular recommendation by treating LLM-cited features as rationales. Rather than taking any single attribution method as ground truth, we evaluate whether selected features carry ranking signal under a common cross-model feature-masking protocol.

\section{Methodology}
\label{sec:methodology}

\subsection{Task Formulation}

We study candidate-level next-basket repurchase recommendation and explanation. Given a customer $c$ and a candidate repurchase item $i$ that appeared in the customer's prior purchase history, we construct a structured feature vector $x_{c,i}$ using only information available before the target basket. The future label $y_{c,i}$ indicates whether item $i$ is purchased in the next basket and is used only for supervised model training, prompt/evidence-card selection, and offline evaluation.

We compare two types of scoring functions. A supervised machine learning ranker produces a score
\begin{equation}
  s_{\mathrm{ML}} = f(x_{c,i}),
\end{equation}
where $f$ is trained on the structured feature set. In parallel, an LLM-based scorer reads a natural-language evidence card $E_{c,i}$ constructed from $x_{c,i}$ and outputs
\begin{equation}
  (s_{\mathrm{LLM}}, R_{\mathrm{LLM}}, z_{\mathrm{LLM}}) = g_{\theta}(E_{c,i}),
\end{equation}
where $s_{\mathrm{LLM}}$ is a behavioral support score, $R_{\mathrm{LLM}}$ is a set of cited top features, and $z_{\mathrm{LLM}}$ is a natural-language explanation. The LLM does not observe the supervised model score $s_{\mathrm{ML}}$ during inference. This design lets us evaluate the LLM both as a standalone feature-based scorer in RQ1 and as a feature-rationale selector in RQ2.

\subsection{Feature Set and Evidence Card}

All models use the same 28 interpretable engineered features computed from purchase history before the target basket. We use structured features rather than raw purchase sequences to provide a controlled and interpretable interface. This allows us to test whether the LLM can reason over explicit behavioral signals without first extracting temporal patterns from long histories, consistent with prior findings that LLM temporal reasoning can be sensitive to how behavioral context is represented~\cite{cao2026context}. The features cover five dimensions: cadence, frequency, recency, user behavior, and item popularity.

For each customer-item pair, we render the feature values as a structured evidence card $E_{c,i}$. Related features are grouped together, and original feature names are retained so that the LLM must cite concrete feature identifiers rather than free-form reasons. The base evidence card contains raw, interpretable feature values. The norm-prompt variant retains these values and additionally provides percentile-based views that contextualize each value relative to an appropriate historical reference distribution without using future labels.

\subsection{LLM Scorer and Rationale Generator}

Given an evidence card, the LLM outputs a repurchase support score between 0 and 1, exactly three feature names, and a natural-language explanation. The LLM does not observe the supervised ranker score or future purchase label during inference. We evaluate its score as a ranking signal in RQ1 and its cited features through the masking analysis in RQ2. Requiring exact feature names links the generated explanation to the structured evidence and prevents rationales based only on generic product semantics.

\subsection{Prompt and Evidence-Card Variants}

We evaluate a base variant and a validation-selected GPT-4o norm-prompt variant. The base variant presents raw feature values and asks the LLM to identify the strongest supporting or weakening evidence.

The norm-prompt uses more neutral feature-selection instructions and augments the raw evidence with percentile-based representations. For customer-specific and customer-item features, each value is converted to its percentile among candidate items for the same customer. For global item-level features, the percentile is computed across the global item population. These representations indicate whether a value is relatively high or low within its appropriate comparison set without using future labels.

Candidate prompt and evidence-card variants are compared on a separate validation set using NDCG@5 computed from the LLM support scores. The selected variant is then fixed for held-out evaluation. Future purchase labels are used only for variant selection and are never included in the evidence card or provided to the LLM during inference. The selected variant's scores and cited features are evaluated under the same protocols as the other LLM methods. A simplified prompt template is provided in Appendix~\ref{app:prompt-template}.

\section{Experimental Setup}
\label{sec:experiments}

\subsection{Datasets and Baselines}

We evaluate on three repurchase recommendation datasets: two public grocery datasets and one proprietary retail dataset. Instacart is a public online grocery dataset for market-basket prediction~\cite{instacart2017}. DC is derived from the Dunnhumby ``Carbo-Loading'' dataset~\cite{dunnhumbycarbo} and provides a complementary public grocery setting with a much smaller product vocabulary. The proprietary dataset is randomly sampled from a large-scale internal grocery platform and serves as an additional production-domain evaluation setting.

\begin{table}[h]
\centering
\caption{Statistics for the two public datasets used in the evaluation. LLM scoring pairs denote the customer-item pairs evaluated by LLM-based scorers and rationale generators.}
\label{tab:dataset_distribution}
\scriptsize
\setlength{\tabcolsep}{4pt}
\begin{tabular}{lrrrr}
\toprule
Dataset & Unique CIDs & Unique Items & Avg. Hist. Items/User & LLM Scoring Pairs \\
\midrule
Instacart   & 100{,}000 & 49{,}071  & 64.6  & 3{,}599 \\
DC          & 283{,}047 & 910      & 9.8   & 1{,}214 \\
\bottomrule
\end{tabular}
\end{table}

Across datasets, candidate items are drawn from each customer's purchase history, features use only information preceding the target basket, and the next basket provides the repurchase labels. Supervised rankers use customer-level data splits to prevent user leakage. Because of LLM inference cost, we sample 100 held-out customers per dataset. Table~\ref{tab:dataset_distribution} summarizes the scale of the two public datasets and their resulting LLM scoring sets.

For RQ1, we compare LLM scores with rankers using the same 28 features. The personal-frequency baseline ranks candidates by \texttt{item\_order\_rate}, while the supervised baselines are XGBoost and a vanilla neural network (VNN), representing tree-based and neural model families. We evaluate Gemini 2.5 Pro, GPT-4o, and the validation-selected GPT-4o norm-prompt variant described in Section~\ref{sec:methodology}. Future labels are used for supervised training and prompt selection but are never provided to the LLM during inference.

\subsection{Evaluation Protocol}

\paragraph{Score utility for RQ1.}
We rank each customer's candidate items by the LLM support score $s_{\mathrm{LLM}}$ and report Precision@$k$, Recall@$k$, and NDCG@$k$. We compare these rankings with the personal-frequency and supervised baselines.

\paragraph{Feature-rationale masking for RQ2.}
For each customer-item pair, a feature selector $A$ returns three features $R_A(c,i)$. We use three features because the LLM is instructed to cite exactly three, ensuring the same masking budget across methods. We compare LLM citations with TreeSHAP features for XGBoost, integrated-gradient features for VNN, and random feature selection.

TreeSHAP evaluated on XGBoost and integrated gradients evaluated on VNN have direct access to the respective target models and therefore serve as target-model-specific reference baselines. The cross-family attribution method provides a comparison baseline without access to the evaluated target model: integrated gradients for the XGBoost target and TreeSHAP for the VNN target. 

We replace each selected feature with its training-set median, producing the perturbed vector $\tilde{x}_{c,i}^{A}$, and re-score it with target model $f$:
\begin{equation}
  \tilde{s}_{f,A}(c,i) = f(\tilde{x}_{c,i}^{A}).
\end{equation}

The masking gap is the resulting degradation in ranking quality:
\begin{equation}
  \Delta_{f,A}@k =
  M@k(f(x)) - M@k(f(\tilde{x}^{A})),
\end{equation}
where $M@k$ is Precision@$k$, Recall@$k$, or NDCG@$k$. A larger gap indicates that the selected features carry more ranking signal for the target model. Evaluating each selector on both XGBoost and VNN tests whether its selected features remain important across model families.

\section{Results}
\label{sec:results}

\subsection{RQ1: LLM Scores as Ranking Signals}

Table~\ref{tab:score-utility} reports ranking performance on the 100-user LLM scoring sets. Across all three datasets, XGBoost and VNN achieve the strongest results, showing that supervised rankers remain substantially more effective than off-the-shelf LLM scorers. 

\begin{table*}[h]
    \caption{Recommendation quality on the 100-user LLM scoring sets for Instacart, DC, and proprietary data. Bold indicates the best value and underline indicates the second-best value in each column across all methods. }
  \label{tab:score-utility}
  \centering
  \scriptsize
  \setlength{\tabcolsep}{3.5pt}
  \begin{tabular}{l l ccc ccc ccc}
    \toprule
    & & \multicolumn{3}{c}{Precision}
    & \multicolumn{3}{c}{Recall}
    & \multicolumn{3}{c}{NDCG} \\
    \cmidrule(lr){3-5}
    \cmidrule(lr){6-8}
    \cmidrule(lr){9-11}
    Type & Method
      & @3 & @5 & @10
      & @3 & @5 & @10
      & @3 & @5 & @10 \\

    \midrule
    \multicolumn{11}{c}{\textbf{Instacart data}} \\
    \midrule
    Heuristic & Personal Top
      & 0.3949 & 0.3435 & 0.3002
      & 0.2652 & 0.3599 & 0.5781
      & 0.6184 & 0.6297 & 0.6481 \\
    ML & XGBoost
      & \textbf{0.4746} & \underline{0.4000} & \underline{0.3209}
      & \textbf{0.3235} & \underline{0.4494} & \underline{0.6330}
      & \textbf{0.6633} & \textbf{0.6962} & \textbf{0.7029} \\
     & Vanilla NN
      & \underline{0.4565} & \textbf{0.4109} & \textbf{0.3220}
      & \underline{0.3121} & \textbf{0.4555} & \textbf{0.6343}
      & \underline{0.6348} & \underline{0.6778} & \underline{0.6965} \\
    LLM & Gemini 2.5 Pro
      & 0.2971 & 0.2630 & 0.2448
      & 0.2126 & 0.2993 & 0.5205
      & 0.4771 & 0.5326 & 0.5607 \\
     & GPT-4o
      & 0.4130 & 0.3783 & 0.3002
      & 0.2682 & 0.4023 & 0.5759
      & 0.6257 & 0.6568 & 0.6853 \\
     & GPT-4o norm-prompt
      & 0.3913 & 0.3522 & 0.3013
      & 0.2575 & 0.3714 & 0.5956
      & 0.6148 & 0.6390 & 0.6640 \\

    \midrule
    \multicolumn{11}{c}{\textbf{DC data}} \\
    \midrule
    Heuristic & Personal Top
      & 0.2645 & 0.2152 & \textbf{0.1650}
      & 0.5761 & 0.7391 & \textbf{0.8478}
      & 0.5662 & 0.6268 & 0.6519 \\
    ML & XGBoost
      & \underline{0.2862} & \underline{0.2196} & \textbf{0.1650}
      & \underline{0.6304} & \underline{0.7609} & \textbf{0.8478}
      & \underline{0.6012} & \underline{0.6528} & \textbf{0.6716} \\
     & Vanilla NN
      & \textbf{0.2935} & \textbf{0.2283} & \underline{0.1629}
      & \textbf{0.6522} & \textbf{0.7935} & \underline{0.8261}
      & \textbf{0.6132} & \textbf{0.6663} & \underline{0.6691} \\
    LLM & Gemini 2.5 Pro
      & 0.1920 & 0.1630 & 0.1207
      & 0.3913 & 0.5217 & 0.8043
      & 0.3659 & 0.4118 & 0.4968 \\
     & GPT-4o
      & 0.2572 & 0.1935 & 0.1607
      & 0.5761 & 0.6413 & 0.8043
      & 0.5331 & 0.5660 & 0.6028 \\
     & GPT-4o norm-prompt
      & 0.2283 & 0.2065 & 0.1609
      & 0.5000 & 0.6848 & 0.8043
      & 0.5085 & 0.5651 & 0.6093 \\

    \midrule
    \multicolumn{11}{c}{\textbf{Proprietary data}} \\
    \midrule
    Heuristic & Personal Top
      & \underline{0.4516} & 0.3828 & 0.2989
      & 0.2857 & 0.3755 & 0.5371
      & 0.6412 & 0.6742 & 0.6846 \\
    ML & XGBoost
      & \textbf{0.4910} & \underline{0.4151} & \underline{0.3108}
      & \underline{0.3091} & \underline{0.4064} & \underline{0.5658}
      & \underline{0.6931} & \underline{0.7162} & \underline{0.7169} \\
     & Vanilla NN
      & \textbf{0.4910} & \textbf{0.4258} & \textbf{0.3215}
      & \textbf{0.3210} & \textbf{0.4298} & \textbf{0.5840}
      & \textbf{0.7127} & \textbf{0.7295} & \textbf{0.7212} \\
    LLM & Gemini 2.5 Pro
      & 0.3728 & 0.3226 & 0.2484
      & 0.2301 & 0.3154 & 0.4574
      & 0.5763 & 0.5953 & 0.6182 \\
     & GPT-4o
      & 0.3978 & 0.3527 & 0.2763
      & 0.2493 & 0.3392 & 0.5086
      & 0.5839 & 0.6239 & 0.6362 \\
     & GPT-4o norm-prompt
      & 0.3118 & 0.3054 & 0.2527
      & 0.1976 & 0.3139 & 0.4632
      & 0.5060 & 0.5600 & 0.5869 \\

    \bottomrule
  \end{tabular}
\end{table*}

GPT-4o nevertheless extracts nontrivial signal from structured behavioral evidence: on Instacart, it outperforms the personal-frequency baseline on most metrics, though it remains below the supervised models. On DC and the proprietary dataset, however, LLM scorers generally trail even the personal-frequency baseline. These results suggest that LLMs can use repurchase features, but do not consistently translate them into effective ranking scores across candidates. The GPT-4o norm-prompt variant also fails to improve ranking performance. Because it changes only the prompt and evidence representation, this result suggests that percentile-based features and prompt refinement alone are insufficient to make an LLM a competitive standalone ranker. The LLM may identify salient signals for individual customer-item pairs while still struggling to translate their relative importance into comparable scores across candidates.

Overall, RQ1 shows that structured behavioral evidence is insufficient for off-the-shelf LLMs to replace supervised next-basket rankers. RQ2 therefore examines whether the features they cite may still provide outcome-grounded explanation evidence independently of their ranking scores.

\subsection{RQ2: Outcome-Grounded Feature Rationales}

\begin{table*}[h]
  \caption{Cross-model feature-masking gaps on the 100-user LLM scoring sets. Larger gaps indicate that the masked top-3 features carried more ranking signal for the target model. Bold indicates the best value and underline indicates the second-best value within each dataset/target-model block.}
  \label{tab:faithfulness}
  \centering
  \scriptsize
  \setlength{\tabcolsep}{2.6pt}
  \begin{tabular}{l l l ccc ccc ccc}
    \toprule
    & & & \multicolumn{3}{c}{Precision Gap}
    & \multicolumn{3}{c}{Recall Gap}
    & \multicolumn{3}{c}{NDCG Gap} \\
    \cmidrule(lr){4-6}
    \cmidrule(lr){7-9}
    \cmidrule(lr){10-12}
    Target & Type & Attribution
      & @3 & @5 & @10
      & @3 & @5 & @10
      & @3 & @5 & @10 \\

    \midrule
    \multicolumn{12}{c}{\textbf{Instacart data}} \\
    \midrule

    XGB & Random & Random top-3
      & 0.0036 & 0.0022 & 0.0054
      & 0.0110 & 0.0339 & 0.0189
      & -0.0017 & -0.0065 & -0.0056 \\
    & ML & XGB SHAP top-3
      & \textbf{0.0507} & \textbf{0.0283} & \textbf{0.0239}
      & \textbf{0.0435} & \textbf{0.0561} & \textbf{0.0648}
      & \underline{0.0413} & \underline{0.0388} & \textbf{0.0307} \\
    &  & VNN IG top-3
      & \textbf{0.0507} & 0.0065 & 0.0141
      & 0.0189 & 0.0133 & \underline{0.0414}
      & 0.0357 & 0.0361 & 0.0109 \\
    & LLM & Gemini-2.5-pro top-3
      & 0.0399 & \underline{0.0130} & 0.0143
      & 0.0216 & 0.0239 & 0.0191
      & 0.0274 & 0.0116 & 0.0119 \\
    &  & GPT-4o top-3
      & \underline{0.0471} & \textbf{0.0283} & \underline{0.0198}
      & 0.0205 & \underline{0.0422} & 0.0408
      & 0.0363 & \textbf{0.0459} & \underline{0.0234} \\
    &  & GPT-4o norm-prompt top-3
      & 0.0435 & 0.0065 & 0.0165
      & \underline{0.0349} & 0.0120 & 0.0369
      & \textbf{0.0484} & 0.0226 & 0.0198 \\

    \addlinespace
    VNN & Random & Random top-3
      & 0.0000 & 0.0152 & 0.0087
      & 0.0046 & 0.0230 & 0.0258
      & -0.0167 & -0.0061 & 0.0001 \\
    & ML & XGB SHAP top-3
      & 0.0507 & 0.0348 & 0.0217
      & \underline{0.0398} & 0.0279 & \underline{0.0488}
      & \underline{0.0366} & \textbf{0.0376} & \textbf{0.0329} \\
    &  & VNN IG top-3
      & \underline{0.0543} & \underline{0.0500} & 0.0163
      & 0.0211 & 0.0378 & 0.0383
      & \textbf{0.0409} & \underline{0.0334} & \underline{0.0245} \\
    & LLM & Gemini-2.5-pro top-3
      & 0.0109 & 0.0348 & \textbf{0.0239}
      & -0.0025 & 0.0309 & 0.0417
      & -0.0152 & -0.0099 & -0.0050 \\
    &  & GPT-4o top-3
      & 0.0399 & \textbf{0.0522} & 0.0141
      & 0.0001 & \textbf{0.0510} & 0.0356
      & 0.0059 & 0.0253 & 0.0158 \\
    &  & GPT-4o norm-prompt top-3
      & \textbf{0.0580} & 0.0413 & \underline{0.0228}
      & \textbf{0.0409} & \underline{0.0446} & \textbf{0.0521}
      & 0.0200 & 0.0144 & 0.0094 \\

    \midrule
    \multicolumn{12}{c}{\textbf{DC data}} \\
    \midrule

    XGB & Random & Random top-3
      & -0.0145 & 0.0000 & 0.0000
      & -0.0435 & 0.0000 & 0.0000
      & -0.0114 & 0.0073 & 0.0100 \\
    & ML & XGB SHAP top-3
      & \underline{0.0217} & \textbf{0.0184} & \textbf{0.0065}
      & \underline{0.0435} & \underline{0.0652} & \textbf{0.0652}
      & 0.0608 & 0.0841 & \underline{0.0835} \\
    &  & VNN IG top-3
      & \textbf{0.0435} & \textbf{0.0172} & 0.0000
      & \textbf{0.1087} & \underline{0.0652} & 0.0217
      & \textbf{0.1317} & \textbf{0.1117} & \textbf{0.0942} \\
    & LLM & Gemini-2.5-pro top-3
      & 0.0072 & 0.0000 & \underline{0.0022}
      & 0.0000 & -0.0109 & 0.0109
      & -0.0057 & -0.0151 & -0.0122 \\
    &  & GPT-4o top-3
      & 0.0000 & -0.0043 & -0.0022
      & -0.0217 & -0.0217 & -0.0109
      & -0.0298 & -0.0323 & -0.0207 \\
    &  & GPT-4o norm-prompt top-3
      & -0.0145 & -0.0087 & 0.0000
      & -0.0543 & -0.0326 & 0.0000
      & -0.0085 & 0.0115 & 0.0144 \\

    \addlinespace
    VNN & Random & Random top-3
      & -0.0072 & 0.0043 & \underline{0.0022}
      & -0.0217 & 0.0109 & 0.0109
      & -0.0120 & 0.0008 & -0.0041 \\
    & ML & XGB SHAP top-3
      & \textbf{0.0652} & \textbf{0.0348} & \textbf{0.0043}
      & \textbf{0.1848} & \textbf{0.1630} & \textbf{0.0435}
      & \textbf{0.1533} & \underline{0.1562} & \underline{0.1055} \\
    &  & VNN IG top-3
      & \underline{0.0362} & \underline{0.0304} & \underline{0.0022}
      & \underline{0.0870} & \underline{0.1196} & \underline{0.0217}
      & 0.1363 & 0.1520 & 0.1047 \\
    & LLM & Gemini-2.5-pro top-3
      & 0.0290 & 0.0043 & 0.0000
      & 0.0761 & 0.0217 & -0.0109
      & 0.0583 & 0.0380 & 0.0208 \\
    &  & GPT-4o top-3
      & 0.0145 & 0.0130 & 0.0000
      & 0.0326 & 0.0652 & 0.0000
      & -0.0017 & 0.0183 & -0.0066 \\
    &  & GPT-4o norm-prompt top-3
      & 0.0145 & 0.0043 & 0.0000
      & 0.0435 & 0.0217 & 0.0000
      & 0.0215 & 0.0131 & 0.0024 \\

    \midrule
    \multicolumn{12}{c}{\textbf{Proprietary data}} \\
    \midrule

    XGB & Random & Random top-3
      & 0.0251 & 0.0151 & 0.0022
      & 0.0159 & 0.0075 & 0.0048
      & 0.0255 & 0.0188 & 0.0229 \\
    & ML & XGB SHAP top-3
      & 0.1219 & \underline{0.0839} & \textbf{0.0387}
      & 0.0828 & \underline{0.0974} & 0.0616
      & \textbf{0.1438} & \underline{0.1258} & 0.0929 \\
    &  & VNN IG top-3
      & \underline{0.1290} & \textbf{0.0968} & \underline{0.0355}
      & \underline{0.0877} & \textbf{0.1148} & \underline{0.0679}
      & \underline{0.1413} & \textbf{0.1470} & \textbf{0.1182} \\
    & LLM & Gemini-2.5-pro top-3
      & 0.1111 & 0.0667 & 0.0290
      & 0.0684 & 0.0508 & 0.0578
      & 0.1042 & 0.0735 & 0.0743 \\
    &  & GPT-4o top-3
      & 0.0394 & 0.0409 & 0.0275
      & 0.0298 & 0.0465 & 0.0351
      & 0.0510 & 0.0589 & 0.0616 \\
    &  & GPT-4o norm-prompt top-3
      & \textbf{0.1398} & \underline{0.0839} & 0.0333
      & \textbf{0.1011} & 0.0830 & \textbf{0.0733}
      & 0.1306 & 0.1015 & \underline{0.0945} \\

    \addlinespace
    VNN & Random & Random top-3
      & 0.0287 & 0.0172 & 0.0065
      & 0.0294 & 0.0254 & 0.0148
      & 0.0324 & 0.0263 & 0.0221 \\
    & ML & XGB SHAP top-3
      & \textbf{0.1075} & \underline{0.0710} & \textbf{0.0344}
      & \underline{0.0811} & \underline{0.0938} & \underline{0.0763}
      & \underline{0.0985} & \underline{0.0840} & 0.0580 \\
    &  & VNN IG top-3
      & \underline{0.0860} & \textbf{0.0817} & \underline{0.0333}
      & \textbf{0.0852} & \textbf{0.1069} & \textbf{0.0805}
      & \textbf{0.1010} & \textbf{0.0921} & \textbf{0.0888} \\
    & LLM & Gemini-2.5-pro top-3
      & 0.0251 & 0.0323 & 0.0140
      & 0.0118 & 0.0387 & 0.0265
      & 0.0171 & 0.0125 & 0.0183 \\
    &  & GPT-4o top-3
      & 0.0287 & 0.0237 & 0.0043
      & 0.0236 & 0.0402 & 0.0185
      & 0.0129 & 0.0243 & 0.0097 \\
    &  & GPT-4o norm-prompt top-3
      & 0.0681 & \underline{0.0710} & 0.0258
      & 0.0547 & 0.0856 & 0.0546
      & 0.0824 & 0.0810 & \underline{0.0701} \\

    \bottomrule
  \end{tabular}
\end{table*}

Table~\ref{tab:faithfulness} reports the cross-model feature-masking gaps, where larger values indicate that the selected top-3 features carry more ranking signal for the target model. As expected, model-specific TreeSHAP and integrated-gradient attributions are generally strong, particularly on the proprietary dataset.

LLM-cited features show a less consistent pattern. Base GPT-4o and Gemini do not reliably match the cross-family attribution baseline, indicating that plausible feature citations are not necessarily outcome-grounded. On Instacart, GPT-4o norm-prompt improves over base GPT-4o on 7 of 18 masking metrics, whereas on the proprietary dataset it improves on all 18 metrics. However, it exceeds the cross-family attribution baseline on 10 of 18 Instacart metrics and only 4 of 18 proprietary metrics. Thus, improvement over the base prompt does not necessarily imply agreement with an attribution baseline. Percentile-based evidence may help the LLM identify relative feature salience, but its effect is dataset-dependent and does not provide a general solution.

DC presents a different pattern. LLM-selected features produce near-zero or negative gaps on the XGBoost target, and norm-prompt improves over base GPT-4o on only 8 of 18 metrics. Although model-specific attribution methods still identify substantial ranking signal, the LLM does not consistently capture it. The smaller product vocabulary and shorter user histories in DC may make the behavioral evidence less discriminative, although our experiments do not isolate the source of this difference. These results reinforce the need to validate LLM-cited rationales within the target domain before deployment.

\paragraph{Text-to-evidence groundedness on Instacart.}
Feature masking evaluates whether cited features carry outcome-relevant ranking signal. As a complementary analysis, we test whether generated explanations accurately express their cited structured evidence. We evaluate 3,599 Instacart explanations, where the LLM shows its clearest ranking signal in RQ1.

We compare the behavioral groups expressed in each explanation with those implied by the cited features and verify that mentioned numerical values match the underlying evidence. Table~\ref{tab:text_groundedness} summarizes the results.

\begin{table}[h]
\centering
\begin{minipage}{0.72\linewidth}
\caption{Text-to-evidence groundedness of generated explanations on the Instacart audit set.}
\label{tab:text_groundedness}
\centering
\scriptsize
\begin{tabular}{lccccc}
\toprule
Dataset & N & Group Prec. & Group Rec. & Group F1 & Numeric Cons. \\
\midrule
Instacart & 3{,}599 & 0.951 & 0.740 & 0.805 & 0.996 \\
\bottomrule
\end{tabular}
\end{minipage}
\end{table}

The explanations achieve high reason-group precision (0.951) and numeric consistency (0.996). The lower recall (0.740) indicates that some cited evidence is omitted from the text, while the high precision indicates that additional reason groups are introduced infrequently. Thus, once features are selected, the LLM generally verbalizes them accurately; selecting features that are outcome-grounded remains the more difficult problem.

\section{Discussion and Limitations}

\paragraph{Repurchase explanations require behavioral evidence.}
Unlike exploratory recommendation, where explanations may rely on item attributes or semantic similarity, repurchase recommendation asks whether a customer is likely to need an item again now. Explanations in this setting should therefore emphasize interpretable behavioral signals such as cadence, frequency, and recency rather than product semantics alone.

\paragraph{Ranking quality and rationale quality can diverge.}
LLM scoring quality and rationale quality do not necessarily move together. GPT-4o norm-prompt does not improve ranking performance, while its gains in outcome-grounded feature selection are uneven across datasets and only inconsistently match or exceed the cross-family attribution baseline. This may reflect the difference between identifying salient evidence for an individual customer-item pair and calibrating comparable scores across many candidates. Feature masking evaluates whether cited features carry predictive signal for trained rankers, but does not establish causal validity or reveal the LLM's internal reasoning. Evaluating only ranking may therefore overlook potential explanation value, while evaluating only textual fluency may overstate the quality of the cited evidence.

\paragraph{Evidence representation matters.}
Norm-prompt changes both the instructions and the representation of evidence by augmenting raw values with within-customer or global percentile views. These relative views may help the LLM identify salient features, but their effects are dataset-dependent and do not improve ranking performance. Prompt and evidence design should therefore be treated as important design choices rather than general solutions.

\paragraph{Toward validated LLM explanation layers.}
Our results support a constrained hybrid design: supervised models rank candidate items, feature-level evaluation identifies outcome-relevant evidence, and LLMs translate that evidence into concise explanations. For example, a cadence-based rationale could state, ``You usually repurchase this item every 10--12 days, and it has been 12 days since your last purchase.'' Such a design uses LLMs as explanation components rather than primary recommendation models. However, our experiments establish only technical grounding conditions; they do not show that the resulting explanations improve user trust or decision quality.

\paragraph{Limitations.}
Our evaluation has several limitations. First, the feature-masking gap measures whether cited features carry predictive signal for trained target models, but it does not establish that these features causally affect future purchases or faithfully represent the LLM's internal reasoning. The masking results should also be interpreted as sensitivity under a particular perturbation scheme rather than absolute feature importance. Replacing selected features with training-set medians may be less natural for binary or categorical variables, while correlated unmasked features may retain redundant information and attenuate the observed gaps.

Second, the text-to-evidence analysis is automatic and limited to Instacart. Although it measures whether generated text reflects the cited groups and numerical values, it does not evaluate whether customers find the explanations clear, useful, persuasive, or trustworthy. Our experiments also do not establish that LLM-based verbalization is necessary or superior to deterministic template-based explanations when the underlying behavioral evidence is already structured.

Third, because LLM inference is costly, we evaluate only 100 customers per dataset. Small differences in ranking or masking metrics should therefore be interpreted cautiously, with greater emphasis placed on consistent patterns across metrics and datasets. The same cost constraint limits the number of prompt and evidence-card variants explored, so the reported norm-prompt should not be viewed as an exhaustive prompt-search result.

Finally, we evaluate two LLM families, two target-ranker families, and a domain-designed set of 28 behavioral features. Other LLMs, ranking architectures, feature sets, or recommendation domains may produce different results. Future work should expand model and domain coverage and evaluate explanation usefulness, user trust, decision speed, and online business impact through controlled user studies or A/B experiments.

\section{Conclusion}

We evaluated LLMs for next-basket repurchase recommendation in two roles: as standalone feature-based scorers and as generators of behavioral feature rationales. Across two public grocery datasets and one proprietary retail dataset, off-the-shelf LLM scores consistently trail supervised rankers. This indicates that providing structured behavioral evidence is not sufficient to make LLMs reliable standalone repurchase recommendation models.

The rationale results reveal a more nuanced role. LLM-cited features sometimes carry measurable outcome-relevant ranking signal, and percentile-based evidence representations can improve feature selection relative to a base prompt in some settings. However, these gains are dataset-dependent and do not consistently match or exceed cross-family attribution baselines.

We also find that, on Instacart, generated explanations closely reflect the cited structured evidence, with high reason-group precision and numeric consistency. Together, these results support a hybrid design in which supervised models rank candidate items, feature-level evaluation validates the cited evidence, and LLMs translate that evidence into concise user-facing rationales. Reliable LLM-based repurchase explanations therefore require evaluating both whether the selected features are relevant to recommendation outcomes and whether the generated text accurately represents those features.

\section*{Declaration on Generative AI}
We used generative AI as part of the evaluated system and for limited writing assistance. In the experimental pipeline, Gemini 2.5 Pro and GPT-4o were evaluated as feature-based scorers and rationale generators, producing recommendation support scores, cited feature rationales, and textual explanations from structured behavioral evidence cards. Generative AI tools were also used for limited language polishing and grammar correction. No generative AI system was used to alter experimental results, tables, or evaluation outcomes. All technical claims, experimental settings, tables, figures, interpretations, and final wording were reviewed and approved by the authors, who take full responsibility for the content of the paper.

\appendix

\section{Prompt Template}
\label{app:prompt-template}

The LLM prompt asks the model to act as a repurchase analyst for a single customer-item pair. Given a structured evidence card, the model is instructed to evaluate cadence, frequency, recency, user context, and item popularity; assign a repurchase support score between 0 and 1; select exactly three feature names from the valid feature set; and generate both a customer-facing explanation and an analyst-facing rationale.

\begin{lstlisting}[caption={Simplified LLM prompt template.}, label={lst:prompt-template}]
You are a repurchase analyst for a grocery delivery service.

Your job is to read a behavioral evidence card for a single (shopper, product) pair and estimate how likely the shopper is to buy that product in their next order.

## Your Task

1. Read the evidence card carefully.
2. Evaluate the available evidence across cadence, frequency, recency, user context, and item popularity.
3. Identify the THREE individual features whose observed values provide the strongest evidence for your final judgment.
   - Select features based on their actual values in this evidence card.
   - Do not select a feature only because its category is generally useful.
   - Features may support a high score, support a low score, or indicate uncertainty.
4. Assign a score of repurchase between 0.0 and 1.0:
   - 0.0 = very unlikely to repurchase next order
   - 0.5 = uncertain or mixed evidence
   - 1.0 = very likely to repurchase next order
5. Write a short customer-facing explanation, 1 to 2 sentences, that:
   - Starts with "You"
   - Uses plain English
   - Mentions the three selected features without exposing raw feature names
   - Does not invent evidence that is missing or marked N/A
6. Write a brief internal reasoning, 1 to 2 sentences, that:
   - Uses exact feature names and their observed values
   - Explains why those features drove the score up or down
   - Is intended for analyst review, not customer display

## Evidence Evaluation Guidance

Treat all evidence dimensions as potentially useful. No dimension is always more important than another. The importance of each feature depends on its value for the current shopper-product pair.

Evaluate each dimension neutrally:

- CADENCE: Does the item appear to follow a regular repurchase rhythm? Is the current timing consistent with, ahead of, or later than that rhythm? If cadence values are N/A, treat cadence as insufficient evidence.
- FREQUENCY: How often has the shopper purchased this item overall and in recent windows? Does the pattern suggest a strong, weak, or occasional habit?
- RECENCY: How recently was the item purchased? Does recent behavior support repurchase, suggest the item was just bought, or indicate that the item has not appeared for a while?
- USER CONTEXT: Does the shopper generally show stable reorder behavior, large or small baskets, frequent orders, or irregular ordering? Use this only if it changes how the item-level evidence should be interpreted.
- ITEM POPULARITY: Does global reorder behavior provide a useful prior for this item? Use this as supporting evidence, not as a substitute for shopper-specific evidence.

When evidence conflicts, reflect the conflict in the score. For example, a product may be frequently purchased but also very recently bought, or globally popular but weak for this specific shopper.

## Feature Selection Rules

When choosing top_features:

- Choose exactly THREE feature names from the valid feature list.
- Use exact Python feature names.
- Select features because their observed values materially affect the score.
- Do not automatically choose one feature from each dimension.
- Do not automatically favor cadence, frequency, or recency.
- Do not cite cadence features when cadence evidence is N/A or insufficient.
- Do not cite item popularity unless it materially affects the judgment beyond shopper-specific evidence.
- Do not cite ID fields such as department_id or aisle_id unless the evidence card explicitly makes them meaningful.
- If two features are redundant, prefer the more interpretable or more direct feature.

## Valid Feature Names for top_features

CADENCE section:
  item_avg_ipi_days             "Avg repurchase interval (days)"
  item_std_ipi_days             "Interval variability (std, days)"
  item_days_overdue_days        "Days overdue / ahead"
  item_avg_ipi_orders           "Avg repurchase interval (orders)"
  item_std_ipi_orders           "Interval variability (std, orders)"
  item_days_overdue_orders      "Orders overdue / ahead"

FREQUENCY section:
  item_purchase_count           "Total times purchased"
  item_order_rate               "Purchase rate (share of all orders)"
  item_order_rate_since_first   "Purchase rate since first buy"
  item_purchase_count_30d       "Purchases in last 30 days"
  item_purchase_count_60d       "Purchases in last 60 days"
  item_purchase_count_90d       "Purchases in last 90 days"

RECENCY section:
  item_orders_since_last        "Orders elapsed since last purchase"
  item_days_since_last          "Days since last purchase"
  item_in_last_order            "Was in most recent order"
  item_streak                   "Current consecutive-order streak"

USER CONTEXT section:
  user_total_orders             "User total prior orders"
  user_total_distinct_items     "User distinct items ever bought"
  user_avg_basket_size          "User avg basket size"
  user_reorder_rate             "User overall reorder rate"
  user_avg_days_between_orders  "User avg days between orders"
  user_std_days_between_orders  "User order-interval variability (days)"

ITEM POPULARITY section:
  product_reorder_rate          "Item global reorder rate"
  product_unique_buyers         "Item unique buyers"
  department_reorder_rate       "Department avg reorder rate"
  aisle_reorder_rate            "Aisle avg reorder rate"
  department_id                 "Department identifier"
  aisle_id                      "Aisle identifier"

## Output Format

Respond with ONLY valid JSON. Do not include markdown or any text outside the JSON object.

{
  "user_id": <integer>,
  "product_id": <integer>,
  "score": <float between 0.0 and 1.0>,
  "top_features": ["<feature_name_1>", "<feature_name_2>", "<feature_name_3>"],
  "explanation": "<customer-facing explanation starting with 'You'>",
  "reasoning": "<analyst-facing reasoning using exact feature names and observed values>"
}

\end{lstlisting}

\end{document}